\documentclass[twocolumn]{emulateapj}

\usepackage{graphicx}
\usepackage{amsmath}
\usepackage{txfonts}
\usepackage{longtable}
\usepackage{lscape}
\usepackage{natbib}
\usepackage{soul}

\makeatletter
\AtBeginDocument{\let\LS@rot\@undefined}
\makeatother

\usepackage[usenames,dvipsnames,svgnames,table]{xcolor}

\shorttitle{Title}%

\shortauthors{Wiehr, Stellmacher \& Bianda}

\begin{document}

\title{Evidence for the two fluid scenario in solar prominences}

\author{E. Wiehr\altaffilmark{1,}, 
G. Stellmacher\altaffilmark{2}, and 
M. Bianda\altaffilmark{3}} 

\altaffiltext{1}{Institut f\"ur Astrophysik, D-37077 G\"ottingen, 
Germany}
\altaffiltext{2}{Institut d'Astrophysique, F-75014, Paris, France}
\altaffiltext{3}{Istituto Ricerche Solari Locarno, associated with Universit\`a della Svizzera italiana, Ch-6605 Locarno, Switzerland}

\email{ewiehr@gwdg.de, stell@iap.fr, mbianda@irsol.ch}

\begin{abstract}

This paper presents observational evidence of the different dynamical 
behavior of neutral and ionized species in solar prominences. The analysis 
of a time-series of Sr\,{\sc ii}\,4078\,\AA{} and Na\,D spectra in a quiescent 
prominence yields systematically larger Doppler shifts (line-of-sight velocities) 
for the ions V$_{LOS}$(Sr\,{\sc ii})= 1.22$\times$V$_{LOS}$(Na\,D). 
Both lines show a 30\,minutes oscillation of good coherence. Sixteen hours later 
the same prominence underwent marked morphological changes (with a rising 
dome), and the Sr\,{\sc ii} velocity excess drops to 
V$_{LOS}$(Sr\,{\sc ii})=1.11$\times$V$_{LOS}$(Na\,D). The same excess is found 
for the line pair Fe\,{\sc ii}\,5018\,\AA{} and He\,{\sc i}\,5015\,\AA{}. The 
widths of the ionic lines, mainly non-thermally broadened, are not related to 
the macro velocities. The emission ratio of Na\,D and Sr\,{\sc ii}, a measure 
of the electron density, yields $n_e = 4\times10^{10}$\,cm$^{-3}$, shows no 
relation with the V$_{LOS}$ variation or with height above the limb, and seems 
to be reduced 16\,hr later during the active phase.
We apply a new wavelength reference from aureola spectra, which is independent 
of photospheric velocity fields.

\end{abstract}


\section{Introduction}
\label{SectIntrod}

The different dynamical behavior of neutral and ionized species has been 
receiving increasing attention recently. It can well be studied in solar prominences  
because, due to their relatively low temperature of T$\le 10^4$ K, they represent a 
partially ionized and weakly collisional plasma with a considerable number of 
neutrals (e.g., Gilbert et al., 2002). Numerical simulations indicate that the 
decoupling of neutral and ionized species in the solar atmosphere can give rise 
to non-ideal MHD effects that can result, for instance, in heating caused by the 
friction between the two species, modify the equilibrium of magnetic structures, 
introduce instabilities under certain magnetic configurations, or change the way 
the various wave modes transform into each other and propagate through the 
atmosphere (see the review by Ballester et al., 2018).
 
Motivated by these considerations different observers tried to measure differences 
between the line-of-sight velocities (LOS) of ions and neutrals. From 
simultaneous time-series spectra, Anan et al. (2017) found no conclusive differences, 
whereas Khomenko et al. (2016) found differences only at restricted locations with 
high velocities in short-lived transients. Stellmacher \& Wiehr (2017), on the 
contrary, found systematically larger LOS velocities for ions (Sr\,{\sc ii}, 
Fe\,{\sc ii}) than for neutral atoms (Na\,{\sc i} and He\,{\sc i}), whereas a
comparison between neutrals (H, He) gives a 1:1 relation. The conspicuous 
similarities between the spectra of Sr\,{\sc ii} 4078 and Na\,D suggest that 
these lines share the same dynamical behavior within a common resolution element 
(in those data about 1.5''$\times$ 1.5''). In order to investigate this 
behavior in more detail, we discuss new observations of the faint metallic 
lines Sr\,{\sc ii}\,4078 and Na\,D. These are optically thin and can be 
expected to respond quite selectively to 'hot' or 'cold' plasma conditions 
because Sr ionizes above $\approx9000$\,K, where neutral Na hardly exists. 

\section{Observations}

From the Locarno observatory (IRSOL) we observed a quiescent prominence 
at the west limb, $5^o$ north, on 2017 June 24-26. Its appearance on June 
24 as a disk filament shows that the prominence is inclined to the limb by 
$\approx45^o$ (Figure\,1). The two emission lines Sr\,{\sc ii}\,4077.71\,\AA{} 
and Na\,D$_2$\,5889.95\,\AA{} are taken in the $13^{th}$ and $9^{th}$ orders, 
respectively, at a fixed grating angle. A simple change of two pre-filters 
separating the orders allows monitoring of both lines at almost the same 
position on the CCD (Figure\,2). 

\begin{figure}[!h]
\centering
\includegraphics[width = 0.47\textwidth]{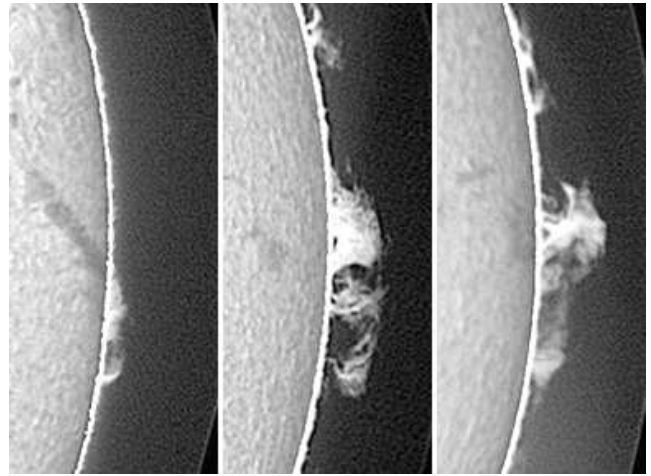}
\caption{Prominence W5N on June 24 (Learmouth; left), 25 (Big Bear; middle),
  26 (Udaipur; right panel); north is up, west is right}
\label{Fig1}
\end{figure}

\begin{figure}[!h]
\centering
\includegraphics[width = 0.49\textwidth]{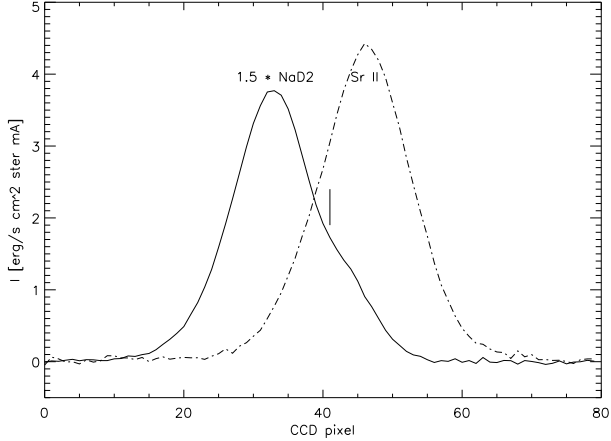}
\caption{Emission lines Na\,D$_2$\,5889.95\,\AA{} ($9^{th}$; 8.4\,m\AA/px) 
and Sr\,{\sc ii}\,4077.71\,\AA{} ($13^{th}$ order; 5.8\,m\AA/px) at a fixed 
grating angle yielding neighboring locations on the CCD; the H$_2$O 5890.7\,\AA{} 
	blend in the red wing of Na\,D$_2$ is indicated by a vertical bar}
\label{Fig2}
\end{figure}

The unchanged grating angle allows measurement of the aureola spectra (giving 
the parasitic light superposing the emission lines) immediately before and
after the prominence spectra. In our former sequential observations
(Stellmacher \& Wiehr, 2017), the change of the grating angle between the 
Sr\,{\sc ii} and the Na\,D exposures required to repeatedly expose aureole 
spectra and thus alternating the telescope pointing between the prominence and 
its (emission-free) neighborhood. This method is sensitive to the pointing 
accuracy and typically requires a time interval of several minutes between 
the observation of both lines.

In the present observations the repetition rate is much shorter due to a 
1.5\,s switch of the two pre-filters. The dimension of the camera chip 
allows coverage of both Na\,D lines. For comparison with former data, we 
additionally observed on June 26 the neighboring emission lines 
He\,{\sc i}\,5015\,\AA{} and Fe\,{\sc ii}\,5018\,\AA{}, and, separately, 
H$_{\delta}$\,4101\,\AA{}. Precise guiding is assured by the accurate 
monitoring of the Gregory telescope's primary image (K\"uveler et al., 2003). 

Disk center spectra are not used for wavelength references but rather to 
calibrate the line intensity in absolute units [erg/(s cm$^2$ ster)], 
using the tables by Labs and Neckels (1970). The spectrograph slit of 
correspondingly 1.5'' width and 120'' length was oriented along the 
direction of refraction (i.e. toward zenith). 

\subsection{Refraction in Earth's atmosphere}

When comparing spectral lines with a marked wavelength difference (here 1818\,\AA{} 
between Sr\,{\sc ii} and Na\,D$_1$), the spectrograph slit must be precisely 
oriented along the direction of refraction to ensure that identical solar 
structures occur in both spectra (shifted perpendicular to the dispersion). 
This is obtained by orienting the slit along the zenith direction, which, however,  
rotates over the solar image. As a consequence, a slit oriented toward the 
zenith sweeps through the solar structures.

\begin{figure}[!ht]
\centering
\includegraphics[width = 0.49\textwidth]{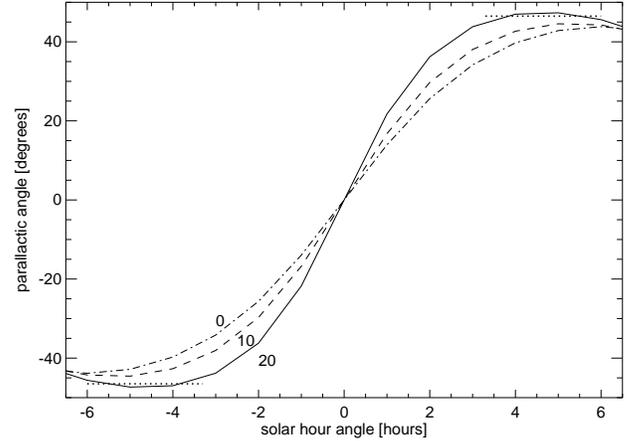}
	\caption{Parallactic angle $\Phi$ as a function of solar hour angle for 
	the Locarno observatory (at $46.17^{\circ}$ latitude) and solar declinations 
	$D_{sun}=0^{\circ}$ (dash-points), D$_{sun}=+10^{\circ}$ (dashes), 
	$D_{sun}=+20^{\circ}$ (full line); observing windows have $\Phi\approx$const. 
	(dotted horizontal lines)}
\label{Fig3}
\end{figure}

The daily variation of the parallactic angle $\Phi$ (spanned by the
geographic north and zenith direction) depends on the solar decl. $D_{sun}$ 
(Figure\,3). For $D_{sun}\ge0^{\circ}$, $\Phi$ shows a minimum and a maximum, 
respectively, at sunrise and at sunset. With increasing $D_{sun}$ these 
extrema become broader and move apart from the sunrise and the sunset. 
For $D_{sun}>20^{\circ}$ (May, 20 through July, 20) and the $46.17^{\circ}$ 
latitude of IRSOL, the two flat extrema allow orientation of the spectrograph 
slit at hour angles $-6.0<\alpha<-3.3$\,hr and $+3.3<\alpha<+6.0$\,hr such, that 
it deviates from the zenith direction by $\le\pm 1°$ (two dotted horizontal lines 
in Figure\,3). This allows taking time-sequence observations up to 2.7\,h with 
an extended slit always covering the same solar structures.  

Balthasar \& Wiehr (1994) made use of the complete morning minimum for 
simultaneous observation of Ca\,{\sc ii}\,8498\,\AA{} and the line pair 
He\,3888\,\AA{} and H$_8$\,3889\,\AA{} in a quiescent prominence.
Anan et a. (2017) observed a similar spectral range of almost 4600\,\AA{} 
from 10:53 to 11:37 local time, where $\Phi$ varies considerably. Here, 
we present observations of Sr\,{\sc ii} and Na\,D in a prominence on 
June 25, during the afternoon maximum and sixteen hours later on June 26, 
during the morning minimum with a slit orientation toward zenith. 

\section{Data reduction}

Prominence emission lines are superposed by parasitic light originating 
from imperfect telescope optics (rather than from Earth's atmosphere; see 
Stellmacher \& Wiehr, 1970). We take spectra of the ''aureola'' in the 
immediate (emission-free) prominence neighborhood with almost the same 
slit inclination to the solar limb. The so obtained aureole spectra are 
normalized to the intensity of the prominence background and then
subtracted (for details see Ramelli et al., 2012). These do not allow full
disappearence of the terrestrial H$_2$O lines by subtraction. (The reason for
this is discussed in Section-3.1.) In particular, remnants of  H$_2$O 5890.7\,\AA\ 
affect the red wing of Na\,D$_2$  (Figure\,2). For comparison with Sr\,{\sc ii} 
we thus prefer Na\,D$_1$, which is not affected by terrestrial H$_2$O lines. 
We verify that both Na\,D lines have the same Doppler shifts, and that their 
integrated line intensities show a fixed relation of D$_2$/D$_1$=1.4 (in 
agreement with Landman, 1981).

We spatially average the spectra over 5 rows (1.7'') adapted to the 
spectrograph slit of correspondingly 1.5'' width. For the spatial 
regions, visually selected to avoid multi-component emissions and marked 
in Figures\,5b and 6b, we determine macro-shifts, V$_{LOS}$, reduced 
widths, $\Delta\lambda_{D}/\lambda_0$, and integrated line intensities, 
$E = I_0 \times \Delta\lambda_{D} \times \pi$ (valid for optically 
thin lines) fitting single Gaussians sequentially to the upper 15\%, 
40\% or 65\% of the central intensity. The wavelengths of the emission 
maxima are determined from the upper 15\%. Reduced widths and 
integrated intensities are only taken from narrow and symmetric line 
profiles, where the Gaussians at 15\%, 40\% or 65\% of the central 
intensity are almost identical. This criterion excludes asymmetric 
profiles; multi-peak profiles are already avoided visually (see above).

\eject
\subsection{Scaling the wavelengths}

As a wavelength reference for Doppler shifts we determine the centers of the 
Sr\,{\sc ii} and Na\,D absorption lines in the aureola at the upper ends of each 
spectrum at 115'' distance from the limb in slit direction (which corresponds 
to 80'' above the equatorial west limb accounting for the slit inclination). 
This reference allows calibration of Doppler shifts independent of the 
complex photospheric velocity fields at disk center and of ubiquitous drifts 
of the spectrograph.

In order to connect these wavelengths from the aureola and those at the solar 
disk, we observe (in 2018 July) aureola spectra of Sr\,{\sc ii}\,4078\,\AA{} 
and Na\,D$_1$ at various distances from the limb. We find that the absorption
lines in the aureola become increasingly blue and red shifted when approaching
the east and west limbs respectively. At the solar poles, these shifts disappear 
(Figure\,4). 

The Doppler shifts in the aureola spectra are found to be equal for the 
Sr\,{\sc ii} and the Na\,D line. Their variation along the slit is the same 
in the aureola and in the emission spectra because the slit inclination is 
largely preserved. The different colors of both lines indeed affect the 
aureole continuum intensity, which is considered by normalization.

\begin{figure}[!h]
\centering
\includegraphics[width = 0.49\textwidth]{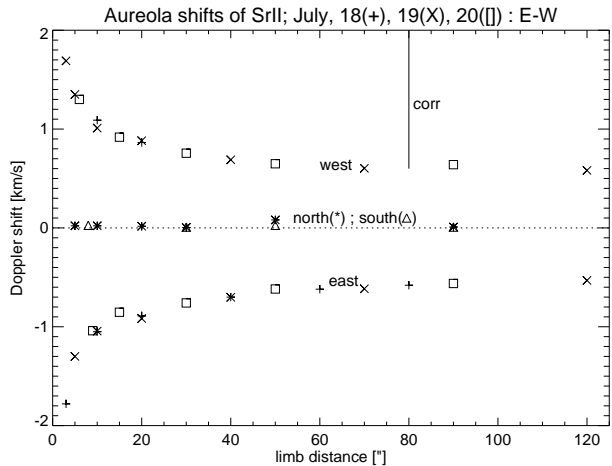}
	\caption{Doppler shift of Sr\,{\sc ii} in the aureola as function of 
	distance from the limb, eastward and westward of the solar equator and 
	northward and southward of the solar poles; the correction to a 
	co-rotating reference is indicated.}
\label{Fig4}
\end{figure}

Figure\,4 indicates that aureola regions closer to the solar disk obtain
parasitic light from increasingly smaller segments of the solar sphere,
which thus imprint increasing rotational Doppler shifts (they are
evidently equal for the Na and the Sr\,{\sc ii} lines) on the respective 
aureola spectra. On the other hand, for increasing limb distances, the 
aureola contains parasitic light from increasingly larger segments, and finally 
from the half-sphere with a much smaller mean rotational Doppler shift. 
Near the poles the sphere segments contain no rotational Doppler shifts, and 
the wavelengths in the aureola then show zero shifts for all limb distances 
(Figure\,4). Hence, the polar aureola gives an almost perfect standard for 
the calibration of Doppler velocities.

\subsection{The co-rotating reference system}

Our wavelengths from the upper end of each spectrum (80'' above the limb) 
may be converted to a co-rotating system. Figure\,4 gives for x=80'' a shift 
of y=0.6\,km/s with respect to the un-shifted polar wavelengths. Assuming 
2\,km/s rotation, our velocity scale had to be shifted by -1.4\,km/s  
(marked in Figure\,4) to relate it to the photosphere below the prominence. 

Since we determine the aureola wavelengths of Sr\,{\sc ii}\,4078 and of
Na\,D at each spectrum of the time series, the resulting macro-velocities
are free from spectrograph drifts and from slow terms of spectrograph 
seeing. The low noise level of the emission lines (see examples in Fig\,2) 
and the Gaussian fit of the upper 15\% lead to an estimated accuracy of 
$\le50$\,m/s for the macro-velocities.

\section{Results}

The H$_{\alpha}$ slit-jaw images from June 25 show that the prominence 
appeared quiet without noticeable morphological changes during the 42.7\,min 
time series (Fig\,5). In contrast to the slit-jaw images, the aspect of the 
spectra (lower panels of Figure\,5) varies at locations with multi-component 
emissions, e.g. at the border of the small emission gap in the prominence center. 
Such multi-components become particularly visible in narrow lines from heavy 
elements ($\mu_{Sr}=86.7$) and are often hidden in broad Balmer lines. We 
select seven spatial scan positions (marked in the lower panels of Figure\,5) 
at regions largely free from multi-components.

\begin{figure}[!h]
\centering
\includegraphics[width = 0.48\textwidth]{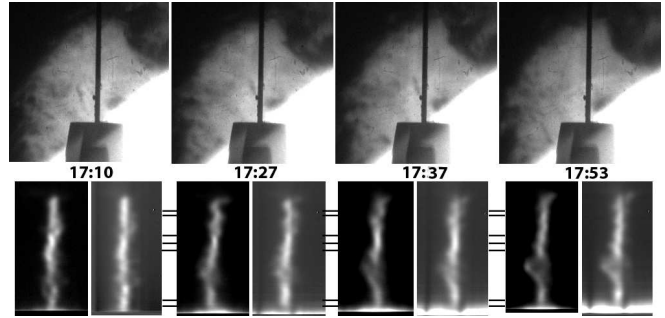}
\caption {Prominence from June 25 at west 5°N. Upper panels: H$_{\alpha}$ 
slit-jaw images at time steps 1, 20, 40, 60; the slit is oriented along the 
direction of refraction. Lower panels: corresponding spectra of Sr\,{\sc ii} 
(left) and Na\,D$_2$ (right of each pair); spatial length = 65'' and the bars 
give the locations of scan rows}
\label{Fig5}
\end{figure}
      
On June 26, 16\,hr after the June 25 observations, the prominence  
is less quiet and characterized by an expansion of the central emission 
gap, which suddenly rose up with a projected motion of $\approx$10\,km/s 
(between 10:04:10 and 10:07:08; see the upper panel of Figure\,6). This value
is close to the one found by Stellmacher \& Wiehr (1973) for a similar event, 
which also showed large Doppler shifts $\le$10\,km/s at the boundary of the
rising cavity. Our spectra even reach 35\,km/s. Yet, Sr\,{\sc ii} and Na\,D
show a striking similarity (see the lower panels of Figure\,6), suggesting
a common origin within the same resolution element, which we estimate to
amount 2''$\times$ 2'', or 1500\,km$\times$1500\,km on the sun.

\begin{figure}[!h]
\centering
\includegraphics[width = 0.48\textwidth]{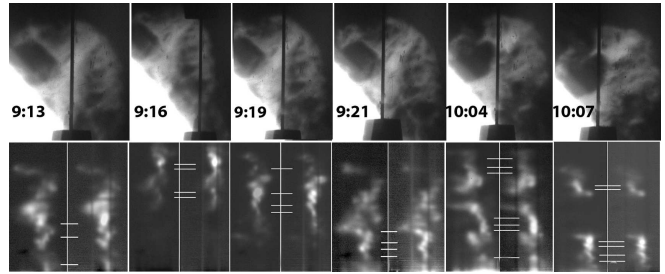}
\caption{Same as in Figure\,5, but 16\,hr later}
\label{Fig6}
\end{figure}

\subsection{Balmer brightness}

H$_{\delta}$ was only observed on June 26. We find a mean integrated intensity 
E$(H_{\delta})=8350$ erg/(s cm$^2$ ster) that, following the tables by Gouttebroze 
et al. (1993) for T=8000\,K, corresponds to $\tau_{\delta}=0.08$. For June 25, we 
estimate the H$_{\delta}$ emission assuming the enhancement of 3.8 with respect 
to June 26, which was measured for Na\,D$_1$, and obtain $\tau_{\delta}=0.15$. 
The H$_{\delta}$ line is thus optically thin on both days. 

For H$_{\alpha}$ the tables give $\tau_{\alpha}=4.0$ and 9.5, respectively, 
and E(H$_{\alpha})=23$ and $38 \times 10^4$ erg/(s cm$^2$ ster). Assuming for 
an elementary volume E(H$_{\alpha})=1 \times 10^4$ erg/(s cm$^2$ster) (see
Stellmacher \& Wiehr, 2000, and references therein) the resolution area covers,
respectively, 23 and 38. Even the larger number is compatible with a
single layer of $\phi\le240$\,km elements in the line of sight and favors the
single line approximation as in Stellmacher \& Wiehr (2017). The spectra show,
indeed, moments where narrower lines coincide with higher line-center intensities,
as expected from the relation $E=I_0 \times \Delta \lambda_D \times \pi$ for 
optically thin lines.

\subsection{Time variation of the Doppler shifts}

\begin{figure}[!h]
\centering
\includegraphics[width = 0.46\textwidth]{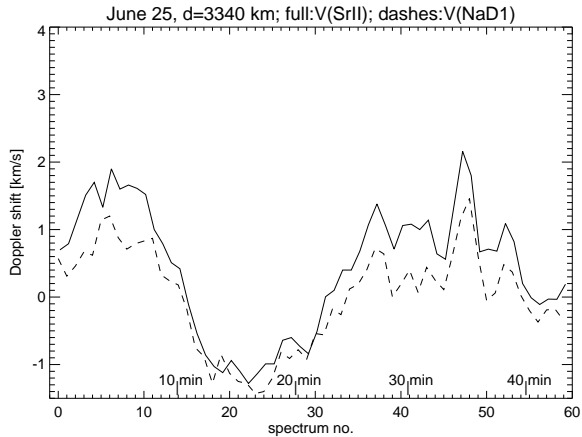}
\caption{Velocity variation of Sr\,{\sc ii} (full line) and Na\,D$_1$ (dashes) 
velocities with respect to the photosphere beneath the prominence for the 60 
steps of the time-series from June 25 in the lowest scan row at d=3340\,km 
from the solar limb in the slit direction}
\label{Fig7}
\end{figure}

\begin{figure}[!h]
\centering
\includegraphics[width = 0.46\textwidth]{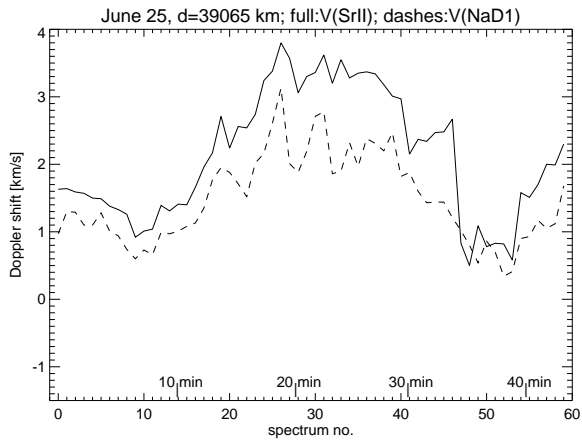}
\caption{Same as Figure\,7 but for the uppermost scan row at d=39065\,km
from the solar limb in the slit direction; the ordinate scales of 
Figures\,7 and 8 are made equal for a direct comparison}
\label{Fig8}
\end{figure}

Figures\,7 and 8 show the V$_{LOS}$ obtained from Sr\,{\sc ii} and Na\,D$_1$ for 
the 60 spectra of the time-series in the lowest and the highest scan rows on 
June 25 (marked in Figure\,5). They are located at 3,860\,km and at 39,600\,km
from the solar limb in slit direction. The time-series (Figures\,7 and 8) show 
wave-like velocity variations with a period of $\approx$30\,min for both lines. 
The V$_{LOS}$ are synchronous in Sr\,{\sc ii} and Na\,D$_1$ with phase shifts 
smaller than the  42.7\,s time step, their amplitudes are systematically higher 
for Sr\,{\sc ii} than for Na\,D$_1$. In Figure\,9 we compare the Sr\,{\sc ii} 
velocities at lower, middle, and upper locations in the prominence on June 25. 
The temporal velocity means indicate an increase from 0.4 via 1.5 to 2.1\,km/s 
through the three levels, whereas the oscillation amplitude remains almost 
constant. This indicates that the oscillation is superposed on an general 
red-shift, which increases with height. The oscillation extrema move along 
the slit direction with 20-40\,km/s.

\begin{figure}[!h]
\centering
\includegraphics[width = 0.46\textwidth]{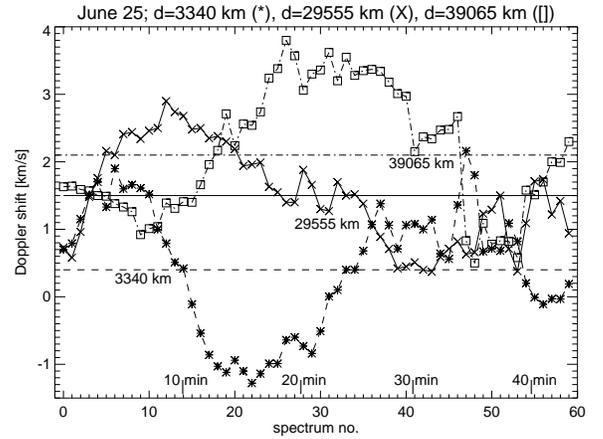}
	\caption{Time variation of Sr\,{\sc ii} velocities in the prominence 
	from June 25, for the 60 spectra at d=3340\,km (dashes), at d=29555\,km 
	(full line) and at d=39065\,km (dashed points) from the solar limb in the 
	slit direction; the ordinate scale for all scans is the same as in Figures\,7 and 8}
\label{Fig9}
\end{figure}

\subsection{Velocity excess of ions}

In Figure\,10 we show the scatter-plot of LOS velocities of Sr\,{\sc ii} 
and Na\,D$_1$ in the co-rotating system (see Sec.\,3.2) for the 60 spectra 
of the time series in the 7 spatial cuts (see Figure\,5). 

\begin{figure}[!h]
\centering
\includegraphics[width = 0.47\textwidth]{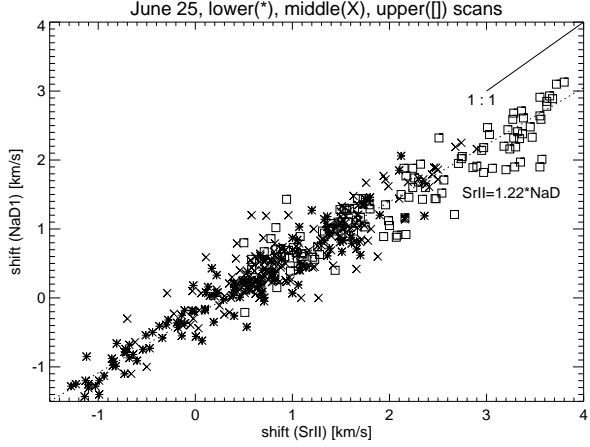}
\caption{Scatter plot of 420 V$_{LOS}$ of Sr\,{\sc ii} and Na\,D$_1$ from the 
7 scan rows and 60 time steps for June 25; the plot shows that the Sr\,{\sc ii} 
velocities are $1.22\pm0.01$ times larger than those of Na\,D}
\label{Fig10}
\end{figure}

From these 420 spectra we obtain a net shift excess of 
V$_{LOS}$(Sr\,{\sc ii})=$1.22\pm0.02\times$ V$_{LOS}$(Na\,D$_1$).  
Concerning the error range, we note that the Gaussian fit to the low-noise 
spectra (see Fig\,2) will not markedly affect the accuracy. We consider
the scatter in Figure\,10 to be due to different influence of image motion 
in the quasi-simultaneous Sr\,{\sc ii} and Na\,D spectra. 

For June 26, we find from 41 scans in the 6 spectra (Figure\,6) velocity excess 
V$_{LOS}$(Sr\,{\sc ii})$=1.10\pm0.03\times$ V$_{LOS}$(Na\,D$_1$) (Figure\,11).
This sample is smaller than on June 25 but contains higher values, notably near 
the rising cavity (see Figure\,6). For the neighboring lines Fe\,{\sc ii}\,5018\,\AA{} 
and He\,{\sc i}\,5015\,\AA{} (singlet line) we find from 27 symmetric profiles a 
velocity excess of V$_{LOS}$(Fe\,{\sc ii})$=1.12\pm0.05\times$ V$_{LOS}$(He\,D$_1$). 
The slight ordinate displacement relative to the Sr\,{\sc ii}-Na\,D 
data (Figure\,11) may arise from the missing reference wavelength of the 
He line, which does not exist in the aureola spectrum and was taken 
from the Fe\,{\sc ii} line via the dispersion. The identical slopes for 
both line pairs in Figure\,11 indicate that the velocity excess of ions 
has actually diminished 16\,hr after the observation of the time-series, 
and that the excess is a standard behavior for ions with respect to neutrals.

\begin{figure}[!h]
\centering
\includegraphics[width = 0.49\textwidth]{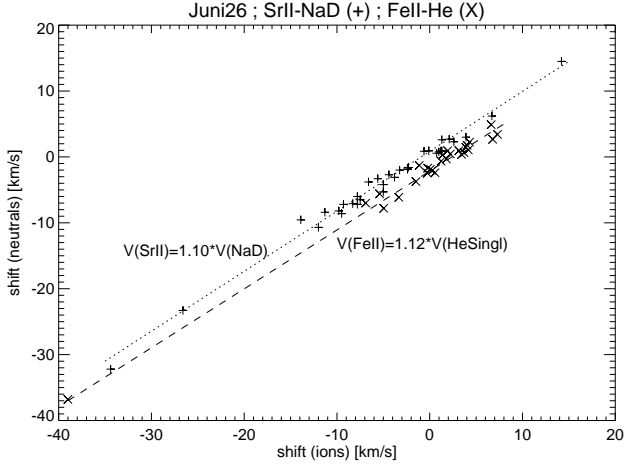}
\caption{Scatter plot of 41 V$_{LOS}$ of Sr\,{\sc ii} and Na\,D$_1$(+) 
  and 27 V$_{LOS}$ of Fe\,{\sc ii}\,5018\,\AA{} and He\,{\sc i}\,5015\,\AA{} (X) 
  for June 26; the ion velocities are $1.11\pm0.05$ times larger than those 
  of the neutrals; the ordinate displacement of V$_{LOS}$(He) reflects the missing 
  reference wavelength in the aureola spectrum}
\label{Fig11}
\end{figure}

\subsection{Integrated line intensities and electron density}

The ratio of total line intensity, $E_{tot}$(Na\,D)/$E_{tot}$(Sr\,{\sc ii}), 
in Figure\,12 does not show the 30\,min period of V$_{LOS}$ (see Figures\,7-9).
It is thus largely independent of macro-shifts. Possible variations of smaller 
periods cannot be established with sufficient significance. In Figure\,13 we plot 
the observed range of emission ratio versus the distance from the solar limb 
in the slit direction and obtain a constant mean of 0.68.

\begin{figure}[!h]
\centering
\includegraphics[width = 0.49\textwidth]{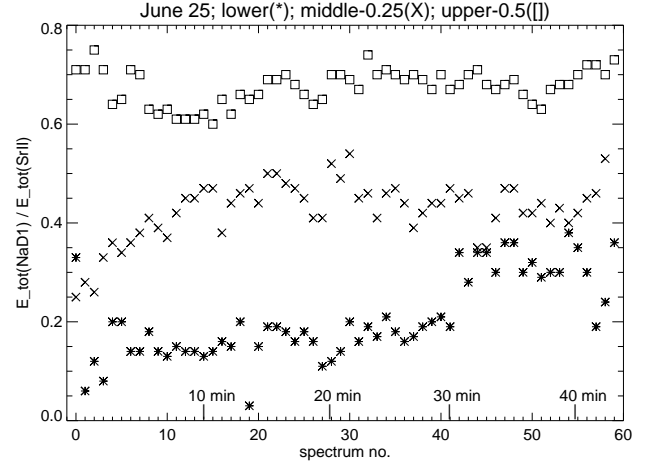}
	\caption{Emission ratio E(Na\,D$_1$) / E(Sr\,{\sc ii}) for the 60 spectra 
	from June 25 at a lower (stars), middle (X-signs), and upper locations (squares) 
	vertically displaced by 0, -0.25, and -0.5 for clearness}
\label{Fig12}
\end{figure}

\begin{figure}[!h]
\centering
\includegraphics[width = 0.49\textwidth]{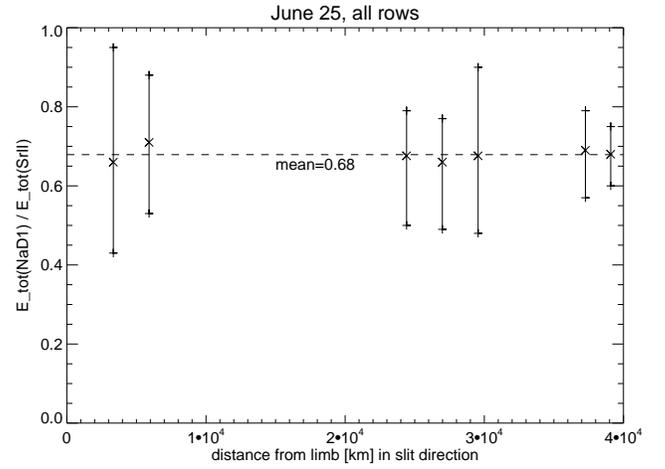}
\caption{Mean emission ratio of Na\,D$_1$ and Sr\,{\sc ii} for all 7 scan rows 
on June 25 as a function of distance from the limb in slit direction}
\label{Fig13}
\end{figure}

\begin{figure}[!h]
\centering
\includegraphics[width = 0.49\textwidth]{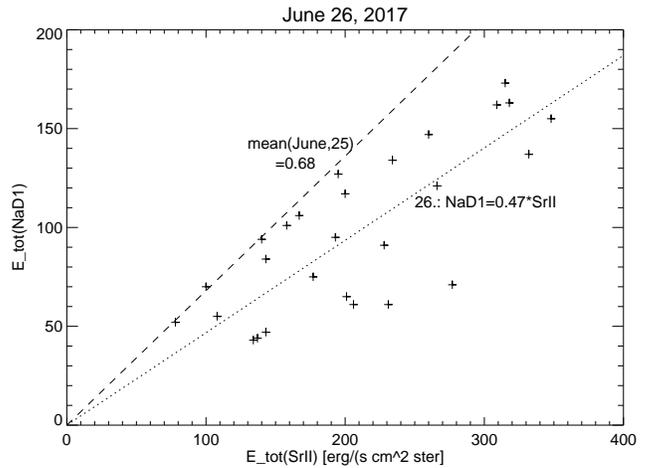}
	\caption{Relation between integrated line intensity of Na\,D$_1$ and 
	Sr\,{\sc ii} for 28 profiles from June 26, at the spatial locations marked
in Figure\,6; all of them are located below the mean relation for June 25
(0.68, dashed line); the mean relation on June 26 is $0.47\pm0.2$}
\label{Fig14}
\end{figure}

The ratio of total line intensity, $E_{tot}$(Na\,D)/$E_{tot}$(Sr\,{\sc ii}),
allows estimation of the electron density. Converting the Na\,D$_1$ emissions 
into Na\,D$_2$ emissions with the factor E(D$_2$)/E(D$_1$)=1.4 (Sec.\,3), we 
obtain from the mean E(Na\,D$_1$)/E(Sr\,{\sc ii})=0.68$\pm0.1$, 
the value E(Na\,D$_2$)/E(Sr\,{\sc ii})=0.95$\pm0.1$ for June 25. Applying 
the calculations by Landman (1983) for T=8000\,K and V$_{nth}= 3$\,km/s 
and the correction factor 0.5 (Landman, 1986), this ratio gives  
n$_e=4\cdot 10^{10}$cm$^{-3}$, which is the same as that found by Stellmacher 
\& Wiehr (2017). 

On June 26, the prominence shows fainter mean emissions; E(Na\,D$_1$) 
reduces by $\approx3.8$, but E(Sr\,{\sc ii}) only by $\approx2.5$. As a 
consequence, the mean emission ratios are generally smaller than those on June 
25. The 28 line profiles, unaffected by multi-component emissions (spatial 
positions marked in the lower panel of Figure\,6), give a wide range of emission 
ratios $0.25\le$ E(Sr\,{\sc ii})/E(Na\,D$_1$)$\le0.7$ (Figure\,14), almost 
entirely below the mean of 0.68 found for June 25. The range of ratios  
gives $1<$n$_e\,[10^{10}$cm$^{-3}]<4$, indicating smaller $n_e$ values 
on June 26, than on June 25. 

\subsection{Width excess of the Sr\,{\sc ii} line}

From symmetric and narrow emissions (i.e. unbiased by multi-component emissions) 
we find mean reduced widths  
$\Delta \lambda_D/\lambda_0(Sr) = 0.95\cdot \Delta \lambda_D/\lambda_0(Na)$. 
The Doppler formula, $\Delta \lambda_D /\lambda_0=1/c\cdot \sqrt{2~R~T_{kin}/
\mu+V_{nth}^2}$, however, leads, for the atomic mass $\mu(Na)=23$ and 
$\mu(Sr)=87.6$, to a markedly smaller ratio of $\sqrt{23/87.6} = 0.51$ thus 
indicating an excess broadening of the Sr\,{\sc ii} line.
$\Delta \lambda_D/\lambda_0$(Sr\,{\sc ii}) does not depend on the 
macro-velocities V$_{LOS}$ (Figure\,15). This is equally found for Na\,D$_1$, 
and also on June 26, even for the larger shifts occurring during the activated 
phase; it agrees with Engvold (1972). The observed range 
$1.0\lesssim\Delta \lambda_D/\lambda_0\,[10^{-5}]\lesssim1.4$ gives 
non-thermal velocities (i.e. for T$_{kin}=0$\,K) of 
$3.0\lesssim V_{nth}\lesssim4.2$ km/s. 

\begin{figure}[!h]
\centering
\includegraphics[width = 0.49\textwidth]{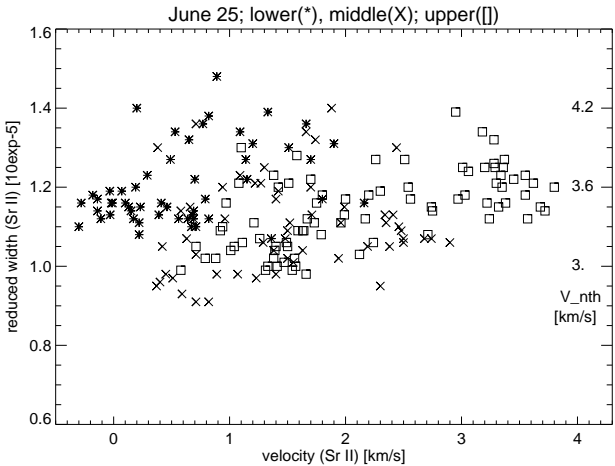}
\caption{Relation between reduced widths and Doppler shifts of Sr\,{\sc ii}
on June 25 for symmetric narrow profiles in the lower, middle, and upper 
scan rows (same as in Figure\,9)}
\label{Fig15}
\end{figure}

\section{Concluding comments}
\label{SectConcl}

\subsection{The systematic velocity excess of ions over neutrals}

The data presented here confirm at higher accuracy the net drift excess of ions 
over neutrals observed by Stellmacher \& Wiehr (2017). This systematic velocity 
excess differs from that of Ca\,{\sc ii}\,8542\,\AA{} over He\,{\sc i}\,10830\,\AA{}, 
which was found by Khomenko et al. (2016) at moments of high velocities in short-lived 
small areas. Our result may in part be due to the judicious choice of the emission 
lines Sr\,{\sc ii}\,4078\,\AA{} and Na\,D, which are optically thin and narrow, 
and to the fact that we analyzed a quiet prominence. 
The smaller velocity excess of 1.11 found for June 26, might be due to a 
a different degree of ionization, as is indicated from the smaller ratio of 
integrated line intensities (see Figure\,14) or to a different density in the 
activated phase of the prominence. 

\eject

\subsection{The electron density}

The estimate of $n_e$ is based on the assumption that Sr\,{\sc ii}\,4078\,\AA{} 
and Na\,D$_2$ originate in the same emission area. The visual aspects of the
spectra (lower panels of Figures\,5 and 6) is in favor of such a common origin.
The small optical thickness ($\tau(H_{\delta})\le0.15$) and the small 
geometric extension of a single layer thickness (see Section\,4) indicate 
a line formation in the same plasma volume. The variation of the emission 
ratio through the observing time (Figure\,12) does not show the 30\,min 
variation of v$_{LOS}$ (see Figure\,7 - 9), indicating that the electron density 
$n_e$ is not related to the macro-velocities. The constancy of $n_e$ with height 
(Figure\,13) is in accordance with our earlier findings (Stellmacher \& Wiehr, 2015, 
2017). For the smaller $n_e$ on June 26, a possible relation between the brightness 
decrease and/or the activation of the prominence on June 26 remains unsolved.  
  
\subsection{The non-thermal line-broadening}

The width excess of the Sr\,{\sc ii} over the Na\,{\sc i} profiles is in 
accordance with Ramelli et al. (2012) and Stellmacher \& Wiehr (2015, 2017) 
who found emission lines from ions to be systematically broader than those 
from neutrals. The conjecture that this width excess may be related to the 
excess of ion velocities seems not be confirmed by the results in 
Figure\,15. Hence, the systematic width excess of lines from ions still 
remains unclear.

\subsection{The oscillatory behavior of the Doppler velocities}

The time-series show a wave-like velocity variation with $\approx30$\,min 
period, which is highly synchronous for ions and neutrals, in agreement with 
Balthasar et al. (1993), Balthasar \& Wiehr (1994), Khomenko et al. (2016) 
and Anan et al. (2017). If we follow that period through the scan rows, it 
seems to travel along the slit direction with a velocity decelerating from 
40\,km/s in the lower to 20\,km/s in the upper scan rows. The superposed 
increase of a mean red-shift (Figure\,9) suggests a swaying motion of the 
prominence as a whole (see Okamoto et al., 2015). Concerning its origin, 
Wedemeyer et al (2013), Hillier et al. (2013) and Wedemeyer \& Steiner (2014) 
showed that the weak prominence magnetic field responds to ubiquitous motions 
of its photospheric footpoints. In this scenario, the motion of ions is 
directly exerted by the Lorentz force, and the motion of the neutrals  
arises from friction to the ions.

\section{Acknowledgments} 
We thank Dr. R. Ramelli (IRSOL) for suggestions with data reduction and fruitful 
discussions, and D. Gisler (IRSOL) for developing the device to shift the pre-filters. 
Drs. R. Hessman (IAP G\"ottingen) and B. Inhester (MPS  G\"ottingen) contributed
helpful comments. E.\,W., G.\,S. thank IRSOL for hospitality. IRSOL is 
supported by the Swiss Confederation (SEFRI), Canton Ticino, the city of Locarno, 
and the local municipalities. We are indebted to the unknown referee for carefully 
reading our manuscript and providing helpful hints to improve it.

\section{References}

Anan T., Ichimoto K., \& Hillier A. 2017, A\&A 601, A103 (DOI 10.1051/0004-6361/201629979)

Ballester, J.-L., Alexeev, I., Collados, M., Downes, T., Pfaff, R.-F., 
Gilbert, H., Khodachenko, M., Khomenko, E., Shaikhislamov, I.-F., Soler, R., 
V\'azquez-Semadeni, E., \& Zaqarashvili, T. 2018, SSRv, 214, 58
(DOI 10.1007/s11214-018-0485-6)

Balthasar, H., Wiehr, E., \& Schleicher, H., W\"ohl, H. 1993, A\&A, 277, 635

Balthasar, H., Wiehr, E. 1994, A\&A, 286, 639

Envold, O. 1972, Solar Phys., 23, 346

Gilbert, H.\,R., Hansteen, V.\,H., Holzer, T.\,E. 202, ApJ, 577, 464 

Gouttebroze, P., Heizel, P., \& Vial, J.-C. 1993, A\&A, Supl. 99, 513

Hillier, A., Morton, R. J., \& Erd\'elyi, R. 2013, ApJ, 779, L16 (DOI 10.1088/2041-8205/779/2/L16)

Khomenko, E., Collados, M., \& Diaz, A.J. 2016, ApJ, 823, 132 (DOI 10.3847/0004-637X/823/2/132)

K\"uveler, G., Wiehr, E., \& Bianda, M.  2003, AN, 324, 308 (DOI 10.1002/asna.200310105)

Labs, D. \& Neckels, H. 1970, Solar Phys., 17, 50

Landman, D. A. 1981, ApJ, 251, 768 

Landman, D. A. 1983, ApJ, 269, 728 

Landman, D. A. 1986, ApJ, 305, 546

Okamoto, T.J., Antolin,P., De Pontieu,B., Uitenbroek, H., Van Doorsselaere, T., \& Yokoyama, T. 2015, ApJ, 809, 71 (DOI 10.1088/0004-637X/809/1/72)
 
Ramelli, R., Stellmcher, G., Wiehr, E., Bianda, M 2012, Solar. Phys. 281, 697 (DOI 10.1007/s11207-012-0118-2)

Stellmacher, G. \& Wiehr, E. 1970, A\&A, 7, 432 

Stellmacher, G. \& Wiehr, E. 1973, A\&A, 24, 321 

Stellmacher, G. \& Wiehr, E. 2000, Solar Phys., 196, 357 (DOI 10.1023/A1005237823016)

Stellmacher, G. \& Wiehr, E. 2015,  A\&A 581, 141 (DOI 10.1051/0004-6361/201322781)

Stellmacher, G. \& Wiehr, E. 2017, Solar Phys., 292, 83 (DOI 10.1007/s11207-017-1103-6)

Wedemeyer, S., Scullion, E., Rouppe van der Voort, L., Bosnjak, A., \& Antolin, P. 2013, ApJ, 774, 123 (DOI 10.1088/0004-637X/774/2/123)

Wedemeyer, S. \& Steiner,O.  2014, PASJ, 66S, 10 (DOI 10.1093/pasj/psu086)

\bibliography{apj-jour,biblio}
\end{document}